\documentclass[10pt]{iopart}

\usepackage{iopams}
\usepackage[T1]{fontenc}
\usepackage[utf8]{inputenc}
\usepackage{textcomp}
\usepackage{float}
\usepackage{graphicx}
\usepackage{tabularx}
\usepackage{subfigure}
\usepackage{grffile}
\usepackage{caption}
\usepackage[dvipsnames]{xcolor}
\usepackage{ulem}
\usepackage{soul}
\usepackage{enumerate}
\usepackage{hyperref}
\usepackage{url}
\usepackage{cite}
\usepackage{todonotes}
\usepackage{multirow}
\usepackage{booktabs}

\graphicspath{{./Figures/}}

\setulcolor{blue}
\setstcolor{red}

\begin{document}

\title{3D modeling of positive streamers in air with inhomogeneous density}

\author{Baohong Guo$^{1}$, Ute Ebert$^{1,2}$, Jannis Teunissen$^{1,*}$}

\address{$^1$ Centrum Wiskunde \& Informatica (CWI), Amsterdam, The Netherlands}
\address{$^2$ Department of Applied Physics, Eindhoven University of Technology, Eindhoven, The Netherlands}

\ead{jannis.teunissen@cwi.nl}

\vspace{10pt}

\begin{indented}
\item[]
\today
\end{indented}

\begin{abstract}
  We study the effect of an inhomogeneous gas density on positive streamer discharges in air using a 3D fluid model with stochastic photoionization, generalizing earlier work with a 2D axisymmetric model by Starikovskiy and Aleksandrov (2019 \textit{Plasma Sources Sci.~Technol.}~\textbf{28} 095022).
  We consider various types of planar and (hemi)spherical gas density gradients.
  Streamers propagate from a region of density $n_\mathrm{0}$ towards a region of higher or lower gas density $n_\mathrm{1}$, where $n_\mathrm{0}$ corresponds to $300\,\mathrm{K}$ and $1\,\mathrm{bar}$.
  We observe that streamers can always propagate into a region with a lower gas density.
  When streamers enter a region with a higher gas density, branching can occur at the density gradient, with branches growing in a flower-like pattern over the gradient surface.
  Depending on the gas density ratio, the gradient width and other factors, narrow branches are able to propagate into the higher-density gas.
  In a planar geometry, we find that such propagation is possible up to a gas density slope of $3.5\,n_\mathrm{0}/\mathrm{mm}$, although this value depends on a number of conditions, such as the gradient angle.
  Surprisingly, a higher applied voltage makes it more difficult for streamers to penetrate into the high-density region, due to an increase of the primary streamer's radius.
\end{abstract}

%
%
%
\ioptwocol

\section{Introduction}\label{sec:intro}

Streamer discharges are an initial phase of electric breakdown when an insulating medium is subjected to a high applied voltage~\cite{nijdam2020a}.
They feature thin plasma channels which largely screen the electric field in their interiors, leading to field enhancement at their tips where they grow due to electron impact ionization.
Streamers are precursors of sparks and lightning leaders~\cite{zhao2022a, hare2019}, and they occur in the upper atmosphere above thunderclouds as sprites~\cite{stenbaek-nielsen2013}, as well as in high-voltage technology~\cite{seeger2017, seeger2018}.
Due to their highly non-equilibrium nature, streamers are also widely used for many plasma applications~\cite{wang2020a, adamovich2022}.

\subsection{Streamer dynamics in varying gas densities}\label{sec:streamer-dynamics}

Streamers can occur at various pressures and in different gases~\cite{briels2008a}.
Streamer dynamics change in different gas densities according to scaling laws~\cite{ebert2010, pasko1998}: all length and time scales of a streamer scale with the inverse gas density $1/n$, the electric field scales as $E \sim n$, and the electron density scales as $n_\mathrm{e} \sim n^2$. 

Streamer propagation in inhomogeneous gas media, where the gas composition or density varies spatially, can be observed under natural and industrial conditions.
This phenomenon has recently been investigated in air by Starikovskiy \textit{et al} using a 2D axisymmetric fluid model~\cite{kosarev2019, starikovskiy2019, starikovskiy2020a}. 
In~\cite{kosarev2019}, the calculations showed how the streamer dynamics changed when a positive streamer developed through a shock wave from a high-density to a low-density region where the air density changed sharply, in agreement with their experimental observations in shock-tube experiments~\cite{kosarev2019}.
In~\cite{starikovskiy2019}, the authors further computationally studied the interaction of streamers with varying air density discontinuities, focusing on the opposite case when a positive streamer propagated from a low-density to a high-density region.  
They found a streamer was unable to penetrate into the high-density region when the density ratio between two different regions was sufficiently large. 
Instead, the streamer was observed to develop along the surface between the two regions.
This phenomenon was summarized by the authors as \textit{``a gas density discontinuity forms a kind of ‘gas-dynamic diode’ that allows the plasma channel to propagate in one direction and blocks its development in another''}~\cite{starikovskiy2019}.
The authors also performed simulations of streamers interacting with gaseous layers of varying densities at both polarities~\cite{starikovskiy2020a}.
They found that negative streamers could pass through a thin low-density layer whereas positive ones could not.

The simulations discussed above were all performed assuming axisymmetric symmetry.
The goal of this paper is to computationally study these phenomena in a full 3D geometry.
This makes it possible to study the effects of streamer branching, which will be shown to be an important mechanism when streamers interact with gas density gradients.
Furthermore, we consider more general gas density gradients than in previous work.


\subsection{Gas density inhomogeneities}\label{sec:density-inhomo}



Gas density inhomogeneities can be induced by shock waves and heating processes.
An early simulation conducted by Marode \textit{et al}~\cite{marode1979} showed that discharge channels could heat air and initiate a radial flow of neutral air molecules, reducing the air density in the path by up to 50\%.
Such air perturbations have been further studied computationally in~\cite{kacem2013, komuro2017a, peng2022}, and measured experimentally in~\cite{ono2004a, ono2010, liu2014, ono2018a, cui2019, komuro2019, komuro2021b, peng2022}.
These studies confirm the presence of spherical or plane shock waves and gas thermal expansion in association with spark and leader discharges.
Such phenomena can heat air to temperatures exceeding thousands of Kelvin and result in a significant decrease in air density.
Furthermore, K{\"o}hn \textit{et al} computationally investigated how sinusoidal air density perturbations induced by discharge shock waves affected streamer properties and the generation of runaway electrons~\cite{kohn2018a, kohn2018b, kohn2020}.

Density perturbations can also be observed around high-speed aircraft or in strong airflows~\cite{grachev1999, fleming2001, gumbel2001, lawson2011, niknezhad2021}.
In plasma aerodynamics, streamer discharges such as nanosecond surface dielectric barrier discharges can be used to control airflow perturbations generated by shock waves, rarefaction waves and jet injection~\cite{bletzinger2005, starikovskii2009, leonov2016, zhang2019c, li2022b}.

Other ways to create gas density inhomogeneities can be the presence of fuel vapor and fuel aerosols or flame in fuel-air mixtures used in plasma-assisted combustion. 
In~\cite{guerra-garcia2015}, experiments were performed in a counterflow non-premixed flame environment, and the results showed that nanosecond pulsed discharges were localized in the area around the front of the counter-flow flame due to a significant decrease in gas density formed by gas heating.

Finally, in Earth's atmosphere the air density varies with altitude, with a scale height of about 8\,km.
This variation affects discharges in the upper atmosphere, in particular so-called sprites, which are essentially streamer discharges.
The effects of gas density on sprites were numerically studied in~\cite{luque2009, luque2010, qin2015}, and experimentally measured (on a much smaller scale) using a hot jet which led to a density ratio of two~\cite{opaits2010}.


\section{3D fluid model}\label{sec:3d-fluid-model}

We simulate positive streamers in dry air consisting of 80\% $\mathrm{N}_2$ and 20\% $\mathrm{O}_2$ at 300\,K with two regions of different air densities connected by a density gradient, see section~\ref{sec:density-gradients}.
Simulations are performed with a 3D drift-diffusion-reaction type fluid model with the local field approximation, using the open-source \texttt{Afivo-streamer} code~\cite{teunissen2017}.

\subsection{Model equations}\label{sec:model-equations}

The electron density $n_{\mathrm e}$ evolves in time as
\begin{equation}\label{eq:evolution-ne}
    \partial_t n_{\mathrm e} = \nabla \cdot (\mu_{\mathrm e} \boldsymbol{\mathrm E} n_{\mathrm e} + D_{\mathrm e} \nabla n_{\mathrm e}) + S_{\mathrm e} + S_{\mathrm{ph}}\,,
\end{equation}
where $\mu_{\mathrm e}$ is the electron mobility, $D_{\mathrm e}$ the electron diffusion coefficient, $\boldsymbol{\mathrm E}$ the electric field.
$S_{\mathrm e}$ is an electron source term due to reactions involving electrons, see table~\ref{tab:chem-reactions}.
$S_{\mathrm{ph}}$ is a non-local photoionization source term described by Zheleznyak's model~\cite{zheleznyak1982}.
This model can be solved by either a continuum (Helmholtz approximation) approach or a stochastic (Monte Carlo) method~\cite{luque2007, bourdon2007, chanrion2008a}, see~\cite{bagheri2019} for a comparison.
We here use stochastic photoionization with discrete ionizing photons using the same parameters as~\cite{wang2023}, which were shown to reproduce streamer branching as observed in experiments~\cite{wang2023}.
For simplicity, we do not take the variation of the photon absorption length with the gas density into account.
The effects of this approximation will be discussed in section~\ref{sec:photoionization-effect}.


All species other than electrons are assumed to be immobile, and we do not consider gas dynamics.
Species densities evolve according to reactions listed in table~\ref{tab:chem-reactions}.
The electric field $\boldsymbol{\mathrm E}$ is calculated as $\boldsymbol{\mathrm E} = - \nabla \phi$. 
The electric potential $\phi$ is obtained by solving Poisson's equation
\begin{equation}
\label{eq:Poisson-equation}
    \nabla^2 \phi = - \rho/\varepsilon_0\,,
\end{equation}
where $\rho$ is the space charge density and $\varepsilon_0$ is the vacuum permittivity.
Equation (\ref{eq:Poisson-equation}) is solved using the geometric multigrid method included in the Afivo library~\cite{teunissen2018, teunissen2023}.

\subsection{Reaction set and input data}\label{sec:reactions-and-input-data}

Since we consider short time scales (up to tens of nanoseconds), we use a relatively small set of chemical reactions, which includes the main reactions between electrons, ions, and neutrals, as shown in table~\ref{tab:chem-reactions}.

\newcounter{nombre}
\renewcommand{\thenombre}{\arabic{nombre}}
\setcounter{nombre}{0}
\newcounter{nombresub} 
\renewcommand{\thenombresub}{\arabic{nombresub}}
\setcounter{nombresub}{0}
\newcommand{\Rnum}[1][]{\refstepcounter{nombre}#1R\thenombre}
\newcommand{\knum}[1][]{\refstepcounter{nombresub}#1$k_{\thenombresub}$($E/n$)}

\begin{table*}
\renewcommand{\arraystretch}{1.1}
\centering
\captionsetup{width=0.95\textwidth}
\caption{List of reactions included in the model, with reaction rate coefficients and references.
The symbol M denotes a neutral molecule (either $\mathrm N_2$ or $\mathrm O_2$).
The reduced electric field $E/n$ is in units of Td (Townsend).
$T$(K) and $T_e$(K) = $2\epsilon_\mathrm{e} / 3k_\mathrm{B}$ are gas and electron temperatures, respectively, where $k_\mathrm{B}$ is the Boltzmann constant and $\epsilon_\mathrm{e}$ is the mean electron energy computed with BOLSIG+~\cite{hagelaar2005}.}
\label{tab:chem-reactions}
\begin{tabular*}{0.95\textwidth}{l@{\extracolsep{\fill}}lll}
 \br
 No. & Reaction & Reaction rate coefficient & Reference \\
 \mr
 \Rnum & $\rm e + N_2 \to e + e + N_2^+$ (15.60\,eV) & \knum & \cite{gradients_phelps_database, phelps1985} \\
 \Rnum & $\rm e + N_2 \to e + e + N_2^+$ (18.80\,eV) & \knum & \cite{gradients_phelps_database, phelps1985} \\
 \Rnum & $\rm e + O_2 \to e + e + O_2^+$ (12.06\,eV) & \knum & \cite{gradients_phelps_database, lawton1978} \\ 
 \Rnum & $\rm e + O_2 + O_2 \to O_2^- + O_2$ & \knum & \cite{gradients_phelps_database, lawton1978} \\
 \Rnum & $\rm e + O_2 \to O^- + O$ & \knum & \cite{gradients_phelps_database, lawton1978} \\ 
 \Rnum & $\rm O_2^- + M \to e + O_2 + M$ & $1.24\times10^{-11}\exp(-(\frac{179}{8.8+E/n})^2)\,\mathrm{cm^3\,s^{-1}}$ & \cite{pancheshnyi2013} \\
 \Rnum & $\rm O^- + N_2 \to e + N_2O$ & $1.16\times10^{-12}\exp(-(\frac{48.9}{11+E/n})^2)\,\mathrm{cm^3\,s^{-1}}$ & \cite{pancheshnyi2013} \\
 \Rnum & $\rm O^- + O_2 + M \to O_3^- + M$ & $1.10\times10^{-30}\exp(-(\frac{E/n}{65})^2)\,\mathrm{cm^6\,s^{-1}}$ & \cite{pancheshnyi2013} \\ 
 \Rnum & $\rm O^- + O_2 \to O_2^- + O$ & $6.96\times10^{-11}\exp(-(\frac{198}{5.6+E/n})^2)\,\mathrm{cm^3\,s^{-1}}$ & \cite{pancheshnyi2013} \\ 
 \Rnum & $\rm N_2^+ + O_2 \to O_2^+ + N_2$ & $6.00\times10^{-11}(\frac{300}{T})^{0.5}\,\mathrm{cm^3\,s^{-1}}$ & \cite{kossyi1992} \\
 \Rnum & $\rm N_2^+ + N_2 + M \to N_4^+ + M$ & $5.00\times10^{-29}(\frac{300}{T})^{2}\,\mathrm{cm^6\,s^{-1}}$ & \cite{kossyi1992, aleksandrov1999} \\
 \Rnum & $\rm N_4^+ + O_2 \to O_2^+ + N_2 + N_2$ & $2.50\times10^{-10}\,\mathrm{cm^3\,s^{-1}}$ & \cite{kossyi1992} \\
 \Rnum & $\rm O_2^+ + O_2 + M \to O_4^+ + M$ & $2.40\times10^{-30}(\frac{300}{T})^{3}\,\mathrm{cm^6\,s^{-1}}$ & \cite{kossyi1992, aleksandrov1999} \\
 \Rnum & $\rm e + N_4^+ \to N_2 + N_2$ & $2.00\times10^{-6}(\frac{300}{T_e})^{0.5}\,\mathrm{cm^3\,s^{-1}}$ & \cite{kossyi1992} \\
 \Rnum & $\rm e + O_4^+ \to O_2 + O_2$ & $1.40\times10^{-6}(\frac{300}{T_e})^{0.5}\,\mathrm{cm^3\,s^{-1}}$ & \cite{kossyi1992} \\
 \br
\end{tabular*}
\end{table*}

Electron transport coefficients $\mu_{\mathrm e}$ and $D_{\mathrm e}$ in equation~(\ref{eq:evolution-ne}), and reaction rate coefficients ($k_1$\,--\,$k_{9}$) in table~\ref{tab:chem-reactions}, are functions of $n$ and of the reduced electric field $E/n$, where $E$ is the electric field and $n$ is the gas number density.
These coefficients were computed with BOLSIG$+$~\cite{hagelaar2005}, with electron-neutral cross sections for $\mathrm N_2$ and $\mathrm O_2$ obtained from the Phelps database~\cite{gradients_phelps_database, lawton1978, phelps1985}.

\subsection{Computational domain and initial condition}\label{sec:com-domain-init-condition}

\begin{figure}
    \centering
    \includegraphics[width=0.48\textwidth]{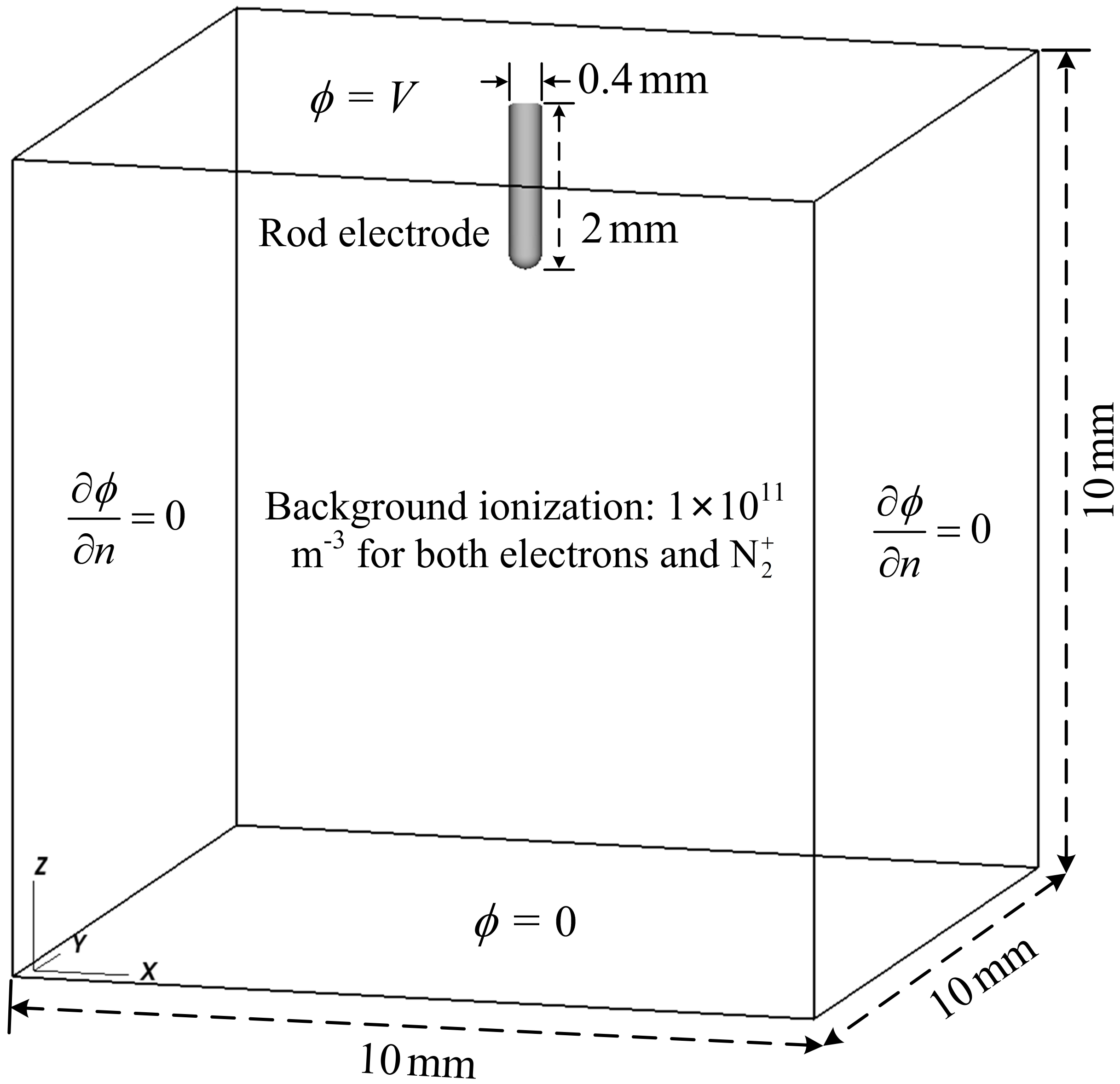}
    \caption{A view of the (10\,mm)$^3$ computational domain.
    The rod electrode protruding from the upper plate from which a streamer starts has a length of 2\,mm and a diameter of 0.4\,mm.
    Boundary conditions for the electric potential $\phi$ are indicated.}
    \label{fig:sim_domain}
\end{figure}

We use a cubic computational domain that measures (10\,mm)$^3$, as illustrated in figure~\ref{fig:sim_domain}.
The domain has a plate-plate geometry with a centrally positioned electrode protruding from the upper plate.
The electrode is rod-shaped with a semi-spherical tip, with a length of 2\,mm and a diameter of 0.4\,mm.

For all species densities, homogeneous Neumann boundary conditions are applied on all domain boundaries, including the rod electrode.
For the electric potential, homogeneous Neumann boundary conditions are applied on the sides of the domain.
The lower plate is grounded, and a constant high voltage $V$ is applied on the upper plate and the rod electrode.
The applied voltage $V$ is always 16\,kV, except for the cases in section~\ref{sec:plane-LTH-voltage} where it is varied between 12\,kV and 18\,kV to investigate its influence on streamer interaction with air density gradients.
In the simulations, the reduced background electric field always remains below the reduced breakdown electric field (where the impact ionization rate is equal to the attachment rate).

As an initial condition, homogeneous background ionization with a density of $10^{11}\,\mathrm m^{-3}$ for both electrons and N$_2^+$ is included for discharge inception.
After inception photoionization will quickly become the dominant source of free electrons for sustaining streamer propagation in air~\cite{nijdam2010}.  
All other ion densities are initially zero.

For computational efficiency, the \texttt{Afivo-streamer} code includes adaptive mesh refinement.
We apply the same refinement criteria to determine the grid spacing $\Delta x$ as~\cite{guo2022d}, and the minimal grid spacing is 1.22\,$\mu$m in the simulations.

\subsection{Air density gradients}\label{sec:density-gradients}

\begin{figure*}
    \centering
    \includegraphics[width=1.0\textwidth]{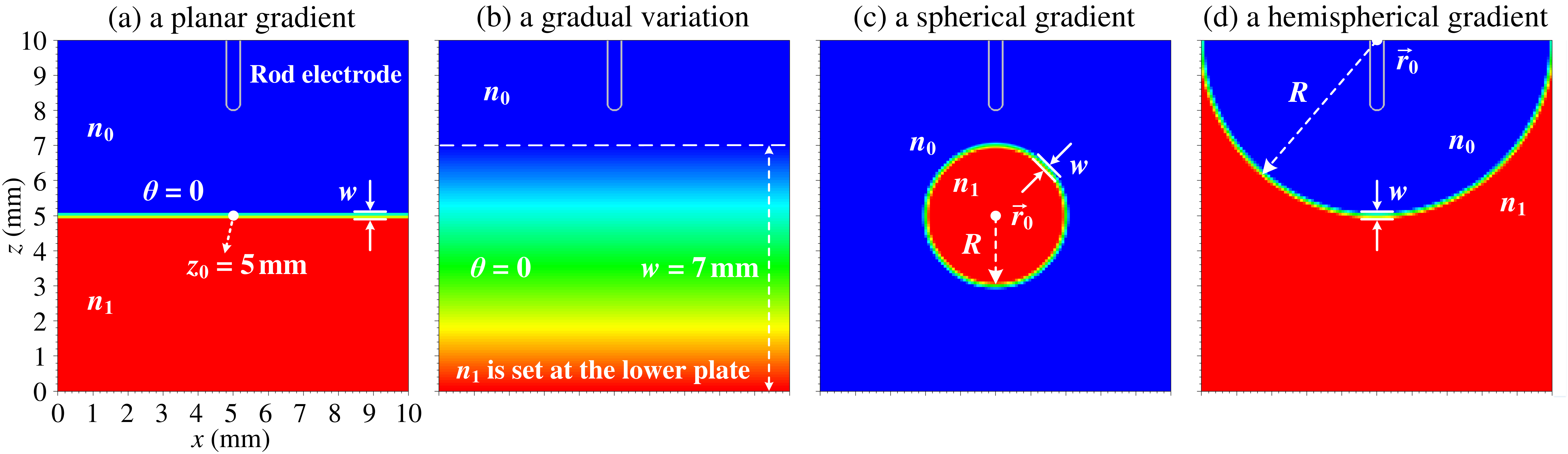}
    \caption{$x$-$z$ view of four types of air density gradients.
    The blue and red regions denote the gas number density as $n_\mathrm{0}$ and $n_\mathrm{1}$, respectively.
    The air density changes linearly in space between these two regions over a width $w$. 
    Panel (a) shows a planar gradient indicating its center height as $z_\mathrm{0}$ and its angle with respect to the $x$-$y$ plane as $\theta$.
    Panel (b) shows a gradual variation with $w=7$\,mm and $\theta=0$.
    Panels (c) and (d) show a spherical and a hemispherical gradient, respectively, where the gradient center is marked as $\vec{r}_{0}$ with a radius $R$.}
    \label{fig:four_gradients}
\end{figure*}

Four types of air density gradients are used, with an example of each illustrated in figure~\ref{fig:four_gradients}.
Two regions can be identified with different gas number densities, denoted as $n_\mathrm{0}$ and $n_\mathrm{1}$, respectively.
Here $n_\mathrm{0}$, located in the region near the rod electrode, is always $2.414\times10^{25}\,\mathrm{m^{-3}}$ (the air density at 1\,bar and 300\,K), whereas $n_\mathrm{1}$ is varied.
For simplicity, we here assume that the temperature is also 300\,K in the region $n_\mathrm{1}$.
Note that in reality, air density differences are due to a combination of temperature and pressure variations, with temperature variations typically persisting longer.
In the simulations, the air density $n$ affects all parameters related to both the density itself and the reduced electric field $E/n$, including reaction rate coefficients ($k_1$\,--\,$k_{9}$) in table~\ref{tab:chem-reactions} and electron transport coefficients $\mu_{\mathrm e}$ and $D_{\mathrm e}$.

The air density changes linearly in space between $n_\mathrm{0}$ and $n_\mathrm{1}$ over a spatial width $w$.
For a planar density gradient shown in figure~\ref{fig:four_gradients}(a), the center height of the gradient is indicated as $z_\mathrm{0}$ and the angle of the gradient with respect to the $x$-$y$ plane is marked as $\theta$.
The gradual density variation shown in figure~\ref{fig:four_gradients}(b) is basically similar to figure~\ref{fig:four_gradients}(a), with a much wider gradient of $w=7$\,mm until the bottom plate.
The spherical and hemispherical density gradients shown in figures~\ref{fig:four_gradients}(c) and (d) are characterized by the gradient center $\vec{r}_{0}$ with a radius $R$.
For details on how these gas density variations were implemented in the code, see~\ref{sec:gas-density-expressions}.
The simulation parameters used in the present paper are summarized in table~\ref{tab:four-gradients}.

\begin{table*}
\centering
\captionsetup{width=1.0\textwidth}
\caption{Summary of four types of air density gradients used in the present paper, corresponding to figure~\ref{fig:four_gradients}.}
\label{tab:four-gradients}
\begin{tabular*}{1.0\textwidth}{l@{\extracolsep{\fill}}llll}
  \br
  Gradient type & planar & gradual & spherical & hemispherical \\
  \mr
  Applied voltages & 12--18\,kV & 16\,kV & 16\,kV & 16\,kV \\
  \multirow{2}{*}{Constant parameters} & \multirow{2}{*}{--} & $w=7$\,mm, & $w=0.2$\,mm, & $w=0.2$\,mm, $R=5$\,mm, \\
  & & $\theta=0$ & $R=2$\,mm & $\vec{r}_{0}=($5\,mm, 5\,mm, 10\,mm) \\
  Variable parameters & $n_\mathrm{1}$, $\theta$, $w$, $z_\mathrm{0}$ & $n_\mathrm{1}$ & $n_\mathrm{1}$, $\vec{r}_{0}$ & $n_\mathrm{1}$ \\
  Simulation results & sections~\ref{sec:plane-gradients} and~\ref{sec:plane-threshold} & section~\ref{sec:gradual-variation} & sections~\ref{sec:spherical-gradient} & section~\ref{sec:hemispherical-gradient} \\ 
  \br
\end{tabular*}
\end{table*}

The time step is restricted according to several criteria as given in~\cite{teunissen2018, teunissen2020}.
Simulations are stopped when the time step becomes smaller than $10^{-14}$\,s or when the discharge reaches the lower plate.
A small time step can occur when a narrow branch forms with a very high electron density and a high electric field, see e.g. figure~\ref{fig:plane_low-to-high_density_ratio}, which leads to a small dielectric relaxation time.
 

\section{Interaction with planar density gradients}\label{sec:plane-gradients}

In this section, we investigate the interaction between a positive streamer and a planar density gradient, see figure~\ref{fig:four_gradients}(a) and table~\ref{tab:four-gradients}.
Both the case where a streamer propagates from a low-density region to a high-density region ($n_\mathrm{1} > n_\mathrm{0}$) and the opposite case ($n_\mathrm{1} < n_\mathrm{0}$) are considered, for which 3D simulation results are presented in sections~\ref{sec:plane-LTH} and~\ref{sec:plane-HTL}, respectively.
In the simulations we vary the center height $z_\mathrm{0}$ of the gradient, the gradient angle $\theta$ (with respect to the $x$-$y$ plane), the gradient width $w$ and the applied voltage $V$ from their default values of $z_\mathrm{0}=5$\,mm, $\theta=0$, $w=0.2$\,mm and $V= 16$\,kV.
The air density ratio $n_\mathrm{1}/n_\mathrm{0}$ is also varied. 

\subsection{From a low-density to a high-density region}\label{sec:plane-LTH}

\subsubsection{Effect of the air density ratio $n_\mathrm{1}/n_\mathrm{0}$}\label{sec:plane-LTH-density-ratio}

\begin{figure*}
    \centering
    \includegraphics[width=1\textwidth]{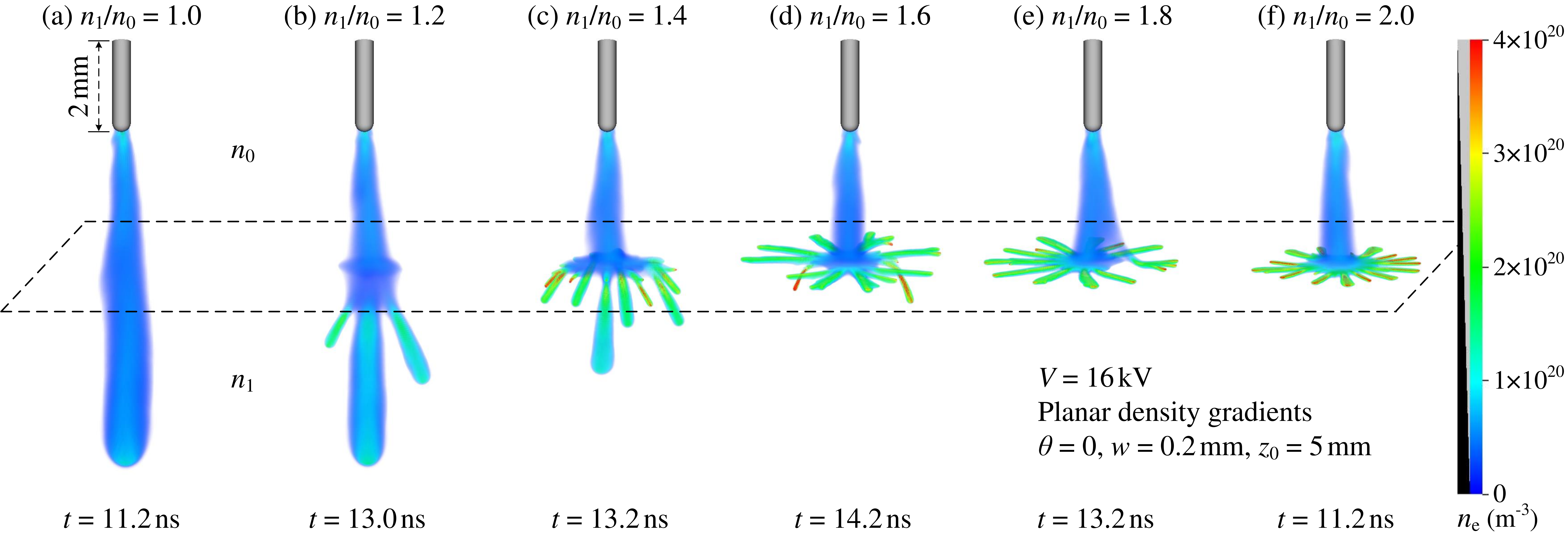}
    \caption{The interaction between a positive streamer in air and a planar density gradient in a 3D simulation as the streamer propagates from a low-density region to a high-density region.
    Streamers for different density ratios $n_\mathrm{1}/n_\mathrm{0}$ at an applied voltage $V=16$\,kV are shown at the last time moment.
    The planar gradient is indicated by a dashed box.
    Shown is a 3D volume rendering of the electron density $n_\mathrm{e}$ with a linear scale ranging from 0 to $4\times10^{20}\,\mathrm{m}^{-3}$ through Visit~\cite{HPV:VisIt}; the opacity is indicated in the legend.
    The same visualization is applied to all subsequent figures presented in the paper.}
    \label{fig:plane_low-to-high_density_ratio}
\end{figure*}

Figure~\ref{fig:plane_low-to-high_density_ratio}(a) shows that a non-branching single streamer is visible in the absence of a density gradient ($n_\mathrm{1}/n_\mathrm{0}=1.0$).
As the density ratio increases to $n_\mathrm{1}/n_\mathrm{0}=1.2$, the planar gradient starts to slightly affect streamer propagation, but the streamer can still penetrate into the high-density region with two extra branches, see figure~\ref{fig:plane_low-to-high_density_ratio}(b).
When $n_\mathrm{1}/n_\mathrm{0}$ further increases to 1.4--1.6, multiple branching channels propagate along the surface of the planar gradient, see figures~\ref{fig:plane_low-to-high_density_ratio}(c) and (d).
The discharge can still enter the high-density region after propagating a certain distance on the surface, which will be further discussed later.
Finally, the streamer is inhibited from propagating through the gradient (up to the time scales considered) when the density ratio is sufficiently high (e.g. $n_\mathrm{1}/n_\mathrm{0} \geqslant 1.8$), see figures~\ref{fig:plane_low-to-high_density_ratio}(e) and (f).
In this case, the streamer propagates along the gradient surface relatively slowly (at a velocity of about 0.2\,--\,0.3$\times10^6$\,m/s) since there is no parallel component of the background electric field along the gradient surface.
The discharge forms a flower-like structure, which is similar to the propagation along a dielectric surface in a barrier discharge.
There are high electric fields at the tips of the branched channels, and the electron density in some channels exceeds $10^{21}\,\mathrm{m}^{-3}$ (which is well above the limit of the color scale).

\begin{figure*}
    \centering
    \includegraphics[width=1\textwidth]{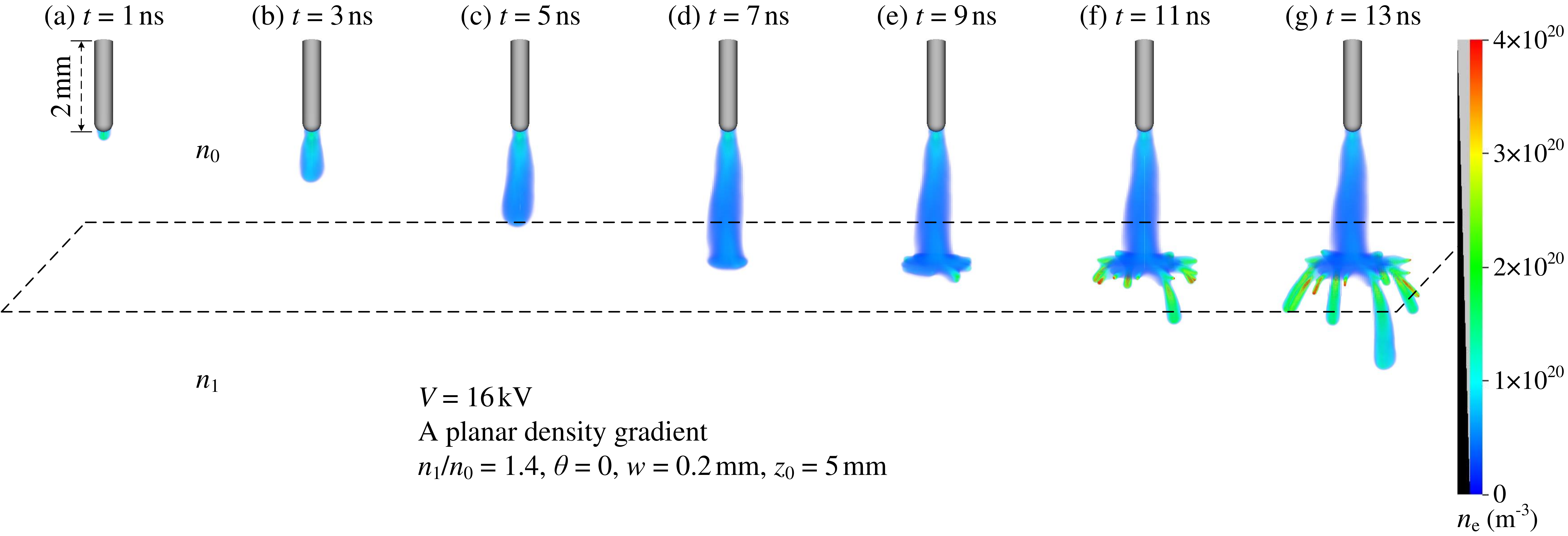}
    \caption{Time evolution of the electron density $n_\mathrm{e}$ for figure~\ref{fig:plane_low-to-high_density_ratio}(c) with $n_\mathrm{1}/n_\mathrm{0}=1.4$ in time steps of 2\,ns.}
    \label{fig:plane_low-to-high_dr1.4_time}
\end{figure*}

The time evolution of the streamer with $n_\mathrm{1}/n_\mathrm{0}=1.4$ in figure~\ref{fig:plane_low-to-high_density_ratio}(c) is shown in figure~\ref{fig:plane_low-to-high_dr1.4_time}.
The streamer initiates from the rod electrode tip and then propagates downwards at a velocity of about $0.5\times10^6$\,m/s, and the streamer radius expands with time until approaching the planar gradient.
When the streamer interacts with the gradient, it is slowed down and inhibited from propagating through the gradient.
Instead, the streamer propagates along the gradient surface at a velocity of about $0.2\times10^6$\,m/s and splits into several branches, leading to decreased radii and increased electric fields at the streamer heads.
Subsequently, this allows the branching channels to eventually enter the high-density region and continue their downward propagation at a velocity of about $0.4\times10^6$\,m/s.
In this region, the streamer channels have smaller radii, higher electric fields, and higher electron densities, as could be expected from the scaling laws mentioned in the introduction.

\subsubsection{Effect of the gradient angle $\theta$}\label{sec:plane-LTH-angle}

\begin{figure*}
    \centering
    \includegraphics[width=1\textwidth]{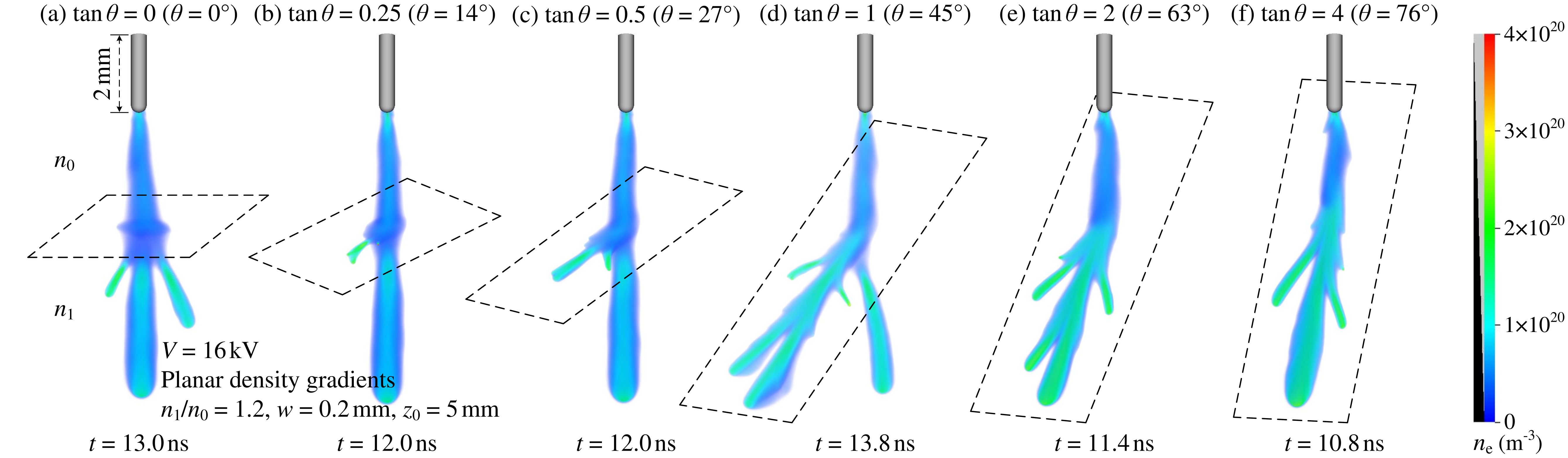}
    \caption{Effect of the gradient angle $\theta$ on streamer propagation for figure~\ref{fig:plane_low-to-high_density_ratio}(b) with $n_\mathrm{1}/n_\mathrm{0}=1.2$.
    The planar gradient is indicated by a dashed box in all panels.
    An increase in $\theta$ facilitates streamer propagation along the gradient surface.}
    \label{fig:plane_low-to-high_angle}
\end{figure*}

The effect of the gradient angle $\theta$ is illustrated in figure~\ref{fig:plane_low-to-high_angle}.
For $\theta=0$ with $n_\mathrm{1}/n_\mathrm{0}=1.2$, the streamer shown in figure~\ref{fig:plane_low-to-high_angle}(a) can propagate through the planar gradient.
Increasing $\theta$ causes the streamer to split into two parts when it encounters the gradient, where the first part penetrates into the high-density region in the original direction, and the second part propagates along the gradient surface, see figures~\ref{fig:plane_low-to-high_angle}(b) and (c).
As $\theta$ further increases to about $\tan \theta=1$ shown in figure~\ref{fig:plane_low-to-high_angle}(d), the surface component becomes dominant, forming several side branches.
When $\theta$ becomes even higher (e.g. $\tan \theta \geqslant 2$) the streamer only propagates along the gradient surface, see figures~\ref{fig:plane_low-to-high_angle}(e) and (f).
In conclusion, increasing the gradient angle $\theta$ facilitates streamer propagation along the gradient surface due to the increased component of the background electric field along the surface.


\subsubsection{Effect of the gradient width $w$}\label{sec:plane-LTH-width}

\begin{figure*}
    \centering
    \includegraphics[width=1\textwidth]{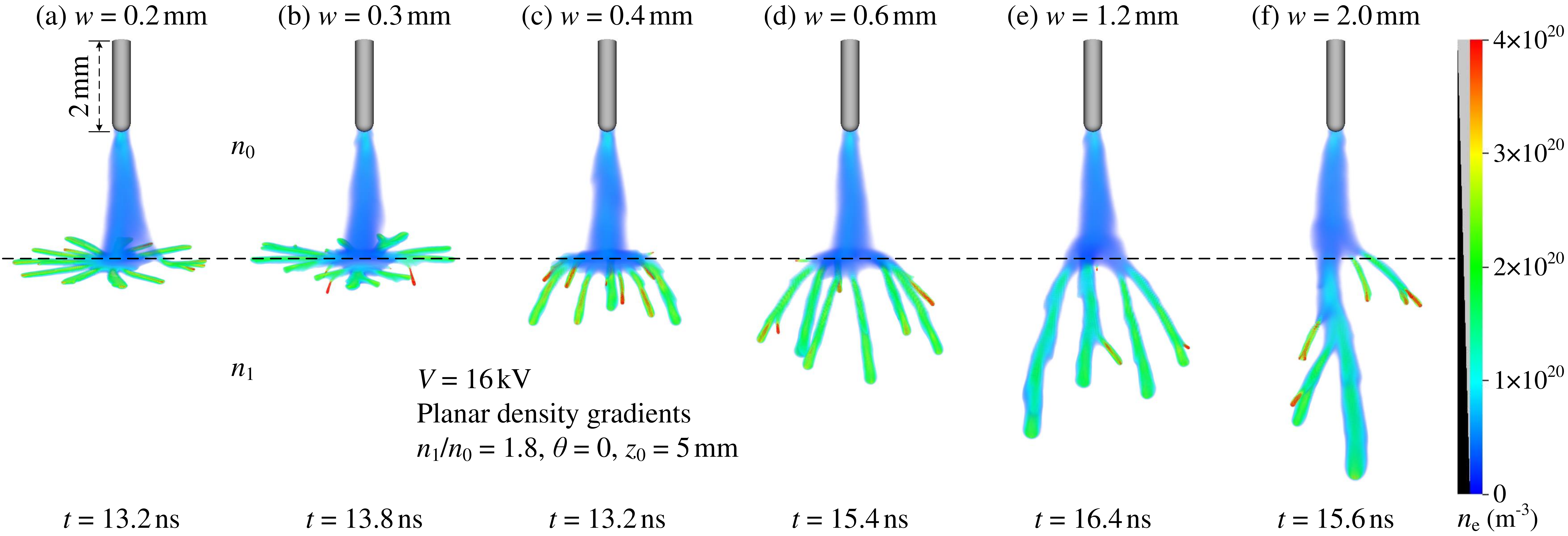}
    \caption{Effect of the gradient width $w$ on streamer propagation for figure~\ref{fig:plane_low-to-high_density_ratio}(e) with $n_\mathrm{1}/n_\mathrm{0}=1.8$.
    The planar gradient is here and afterward indicated by a dashed line.
    An increase in $w$ allows the streamer to overcome the gradient.}
    \label{fig:plane_low-to-high_width}
\end{figure*}

Figure~\ref{fig:plane_low-to-high_width} illustrates the effect of the gradient width $w$.
For $w=0.2$\,mm with $n_\mathrm{1}/n_\mathrm{0}=1.8$ and $\theta=0$, the streamer is inhibited from propagating through the gradient and it forms a flower-like structure, see figure~\ref{fig:plane_low-to-high_width}(a).
However, this inhibition can be overcome by increasing the gradient width.
When $w$ increases to 0.3\,mm shown in figure~\ref{fig:plane_low-to-high_width}(b), most of the branching channels still propagate along the gradient surface, whereas the rest of them enter the high-density region with much higher electron densities inside the thin branching channels.
For $w$ exceeding 0.4\,mm, all streamer branches can eventually penetrate into the high-density region, with fewer branching channels being formed for larger $w$, see figures~\ref{fig:plane_low-to-high_width}(c)--(f).
Note that Starikovskiy \textit{et al} observed a similar effect that allowed the streamer to propagate through the gradient by increasing the gradient width in~\cite{starikovskiy2019}, which occurred when the gradient width became comparable to the streamer radius.
Here we find that this effect also depends on the density slope, which will be further elaborated in section~\ref{sec:plane-threshold}.

\subsubsection{Effect of the applied voltage $V$}\label{sec:plane-LTH-voltage}

\begin{figure*}
    \centering
    \includegraphics[width=0.75\textwidth]{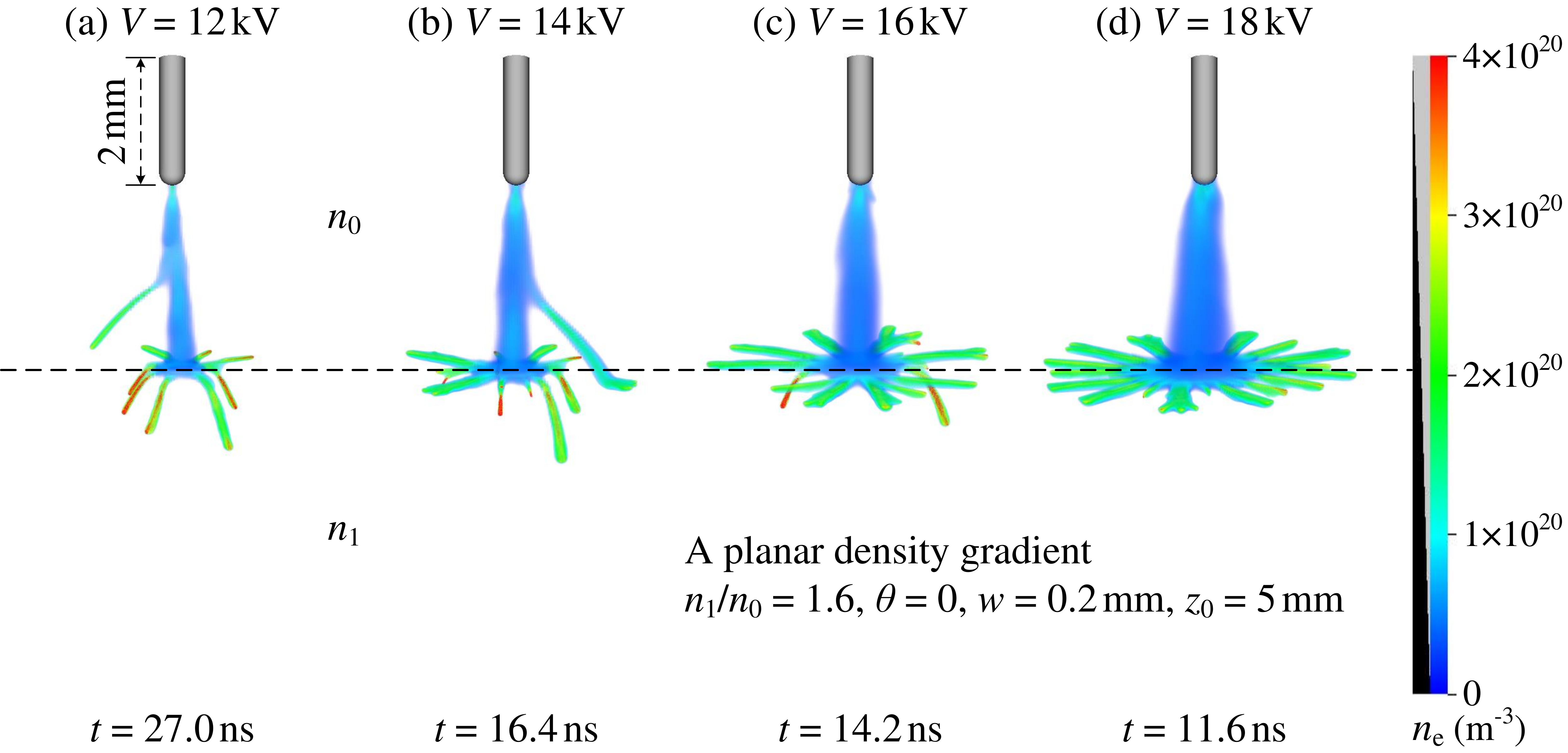}
    \caption{Effect of the applied voltage $V$ on streamer propagation for figure~\ref{fig:plane_low-to-high_density_ratio}(d) with $n_\mathrm{1}/n_\mathrm{0}=1.6$.
    Surprisingly, the streamer more easily propagates through the gradient with a lower applied voltage.} 
    \label{fig:plane_low-to-high_voltage}
\end{figure*}

The effect of the applied voltage $V$ is illustrated in figure~\ref{fig:plane_low-to-high_voltage}.
For $V=12$\,kV with $n_\mathrm{1}/n_\mathrm{0}=1.6$, $\theta=0$ and $w=0.2$\,mm, the streamer can propagate through the gradient and form several branching channels, see figure~\ref{fig:plane_low-to-high_voltage}(a).
As $V$ increases to 14--16\,kV, the radius of the streamer in the low-density region also increases.
This makes it more difficult for the streamer to enter the high-density region, resulting in the formation of more branching channels along the surface, see figures~\ref{fig:plane_low-to-high_voltage}(b) and (c).
When $V$ further increases to 18\,kV, the streamer is inhibited from propagating through the gradient, see figure~\ref{fig:plane_low-to-high_voltage}(d). 
Note that the interaction between a positive streamer and a planar gradient resembles the behavior of archery, where an arrow with a sharper tip is more likely to pierce the target.
In addition, the streamer in the low-density region is more prone to branch at lower applied voltages.

In~\cite{starikovskiy2019} it was argued that the ability of a streamer to propagate through a sharp density discontinuity increased with the applied voltage due to an increase in the reduced background electric field $E/n$.
However, in figure~\ref{fig:plane_low-to-high_voltage} we observe the opposite effect: a streamer more easily propagates through a planar gradient with a lower applied voltage.
Whether a streamer can propagate through such a gradient depends on many factors, including e.g. the streamer radius, the background electric field and the gradient angle.
For the cases considered here the main factor appears to be the streamer radius, with a larger radius making it more likely for the streamer to deform into a ``surface'' discharge.

\begin{figure*}
    \centering
    \includegraphics[width=1\textwidth]{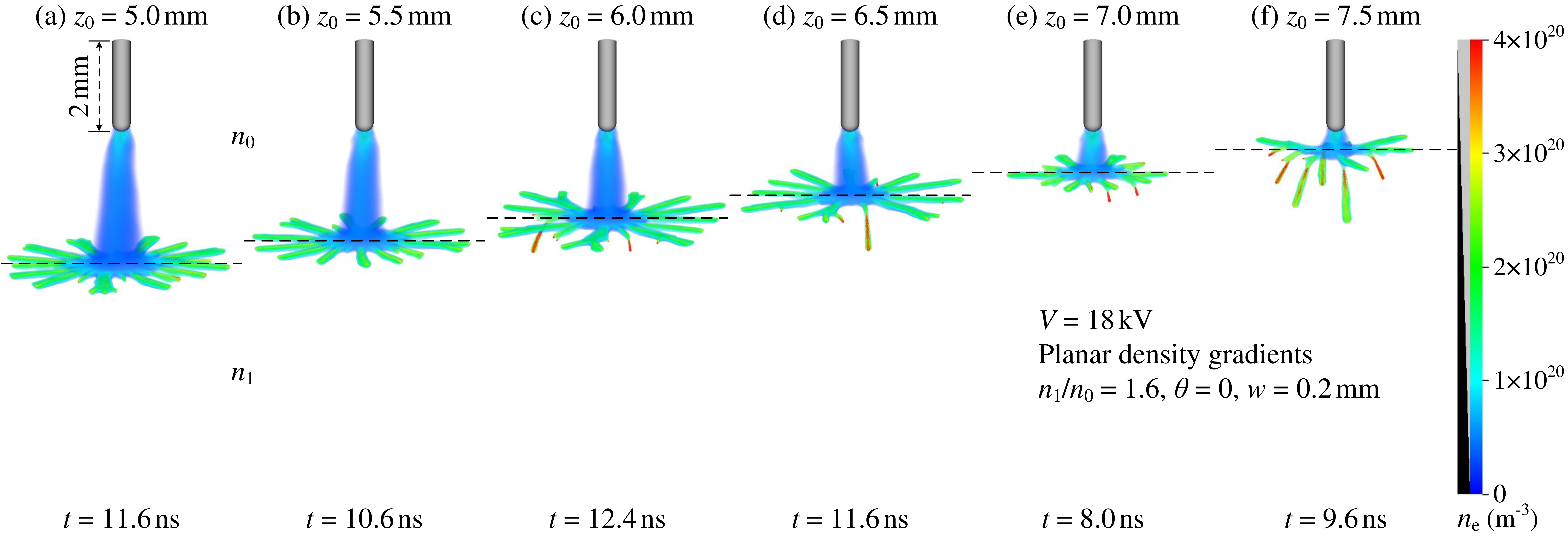}
    \caption{Effect of the distance between the electrode tip and the gradient on streamer propagation.
    The center height $z_\mathrm{0}$ of the gradient is varied, and other parameters are the same as in figure~\ref{fig:plane_low-to-high_voltage}(d).
    When the gradient is close to the electrode ($z_\mathrm{0} \geqslant 6.0$\,mm), the streamer can propagate into the high-density region.}
    \label{fig:plane_low-to-high_1.8e6_height}
\end{figure*}

In figure~\ref{fig:plane_low-to-high_1.8e6_height} we vary the center height $z_\mathrm{0}$ of the gradient, with other parameters the same as in figure~\ref{fig:plane_low-to-high_voltage}(d).
As expected, reducing the distance between the electrode tip and the gradient results in a smaller streamer radius and an increased background field at the gradient, thereby collectively facilitating the propagation of the streamer into the high-density region.

\subsection{From a high-density to a low-density region}\label{sec:plane-HTL}

\begin{figure*}
    \centering
    \includegraphics[width=1\textwidth]{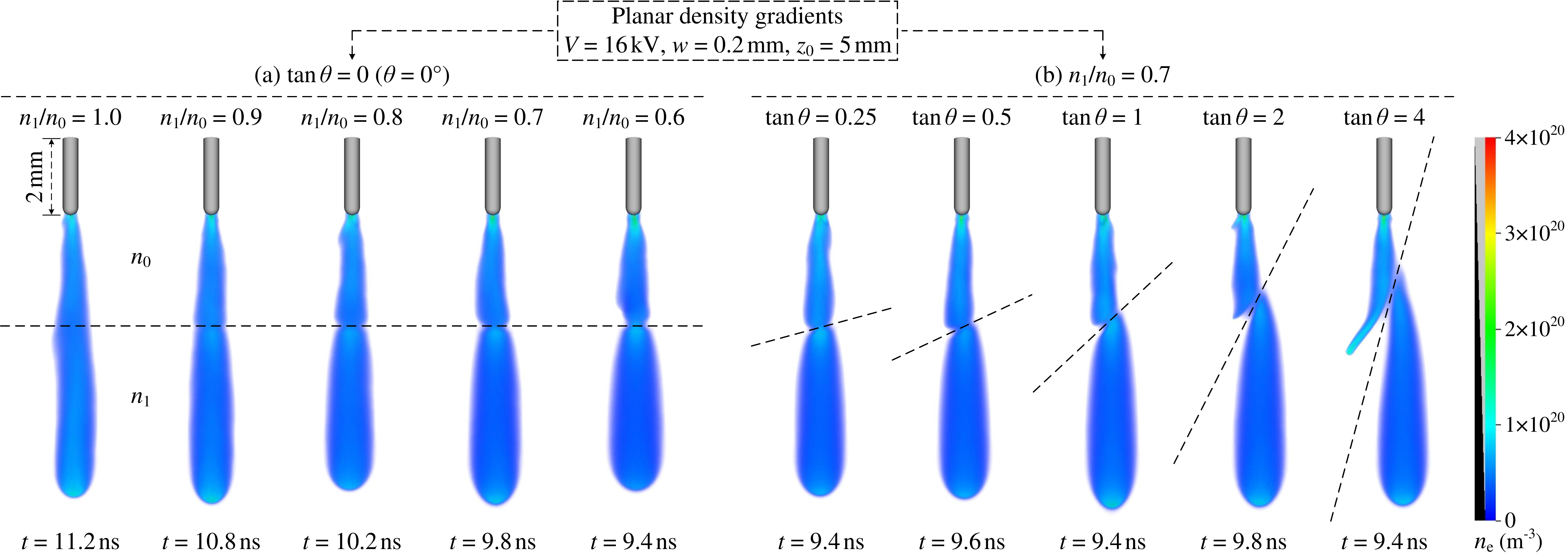}
    \caption{Streamer propagation from a high-density region to a low-density region.
    Panel (a) shows streamers for different density ratios $n_\mathrm{1}/n_\mathrm{0}$ at an angle $\theta=0$, whereas panel (b) shows streamers for different gradient angles $\theta$ at a density ratio $n_\mathrm{1}/n_\mathrm{0}=0.7$.}
    \label{fig:plane_high-to-low}
\end{figure*}

Figure~\ref{fig:plane_high-to-low} shows the effects of the air density ratio $n_\mathrm{1}/n_\mathrm{0}$ and the gradient angle $\theta$ on streamer propagation.
As expected, for $n_\mathrm{1}/n_\mathrm{0} < 1$ with $\theta=0$ the streamer can propagate through the planar gradient, forming a small discontinuity at the gradient, see figure~\ref{fig:plane_high-to-low}(a).
This discontinuity of streamer propagation is attributed to the change in the reduced background electric field $E/n$ caused by the air density gradient.
In the low-density region the streamer radius increases, whereas the electric field at the streamer head and the electron density in the streamer channel decrease compared to the high-density region.


The effect of the gradient angle $\theta$ is illustrated in figure~\ref{fig:plane_high-to-low}(b).
For $\theta > 0$ with $n_\mathrm{1}/n_\mathrm{0}=0.7$, the streamer deviates from its original path after it encounters the gradient. 
More specifically, the streamer propagates perpendicular to the gradient upon encountering it, and continues its propagation along the background electric field in the low-density region.
This deviation becomes more prominent as $\theta$ increases.
In addition to the primary channel, another branching channel initiated by the lower tip of the streamer when encountering the gradient is observed to continue propagating in the high-density region when $\theta$ increases to $\tan \theta=4$.
Note that this branching channel is repelled from the gradient surface by the primary channel.

\section{Interaction with non-planar gradients}\label{sec:non-planar-gradients}

\subsection{Spherical density gradient}\label{sec:spherical-gradient}

We first investigate how a positive streamer interacts with a spherical density gradient, see figure~\ref{fig:four_gradients}(c) and table~\ref{tab:four-gradients}. 
Similar to section~\ref{sec:plane-gradients}, we consider both cases where $n_\mathrm{1} > n_\mathrm{0}$ and $n_\mathrm{1} < n_\mathrm{0}$.
In the simulations, the air density ratio $n_\mathrm{1}/n_\mathrm{0}$ and the gradient center $\vec{r}_{0}$ are varied, whereas the sphere radius $R=2$\,mm, the gradient width $w=0.2$\,mm and the applied voltage $V=16$\,kV are kept constant.

\begin{figure*}
    \centering
    \includegraphics[width=1\textwidth]{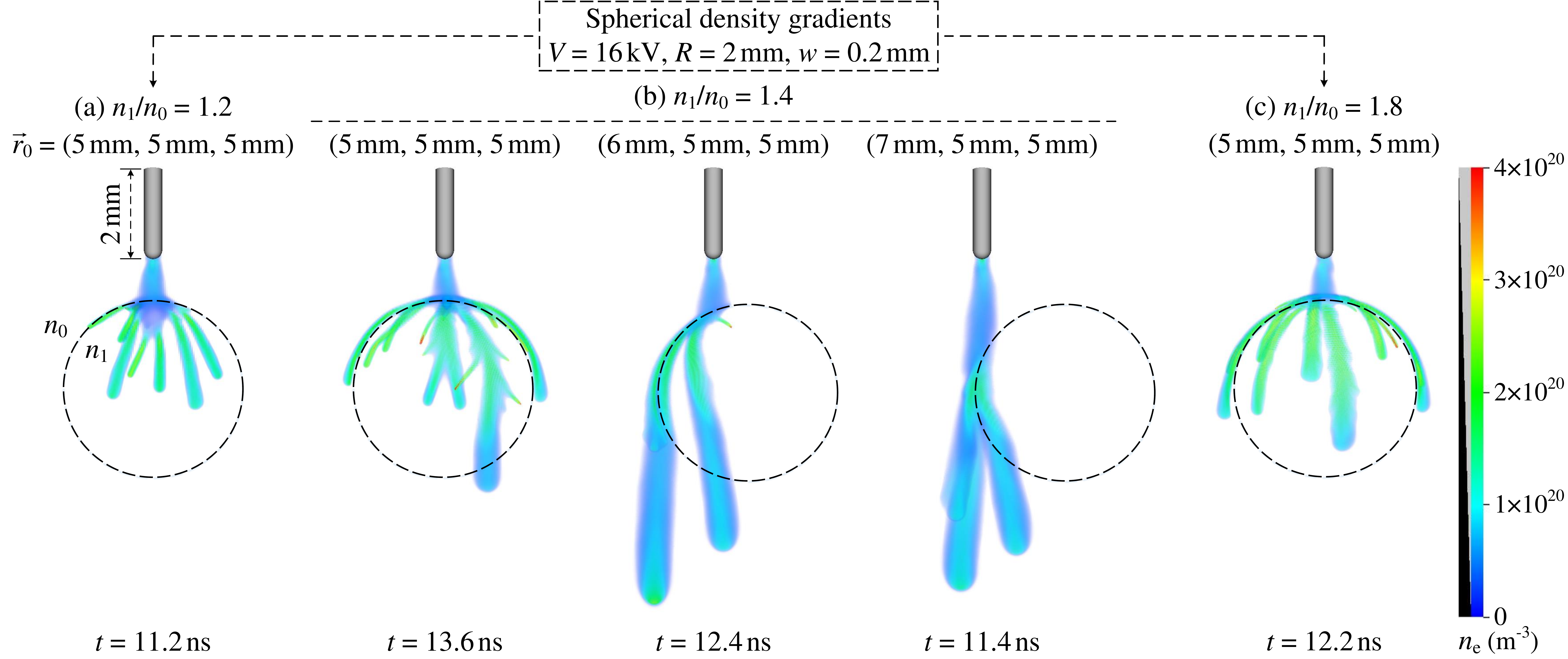}
    \caption{The interaction between a streamer and a spherical density gradient for $n_\mathrm{1}/n_\mathrm{0} > 1$.
    Panels (a) and (c) show streamers for different density ratios $n_\mathrm{1}/n_\mathrm{0}$ at a gradient center $\vec{r}_{0}=($5\,mm, 5\,mm, 5\,mm), whereas panel (b) shows streamers for different gradient centers $\vec{r}_{0}$ at a density ratio $n_\mathrm{1}/n_\mathrm{0}=1.4$.
    The spherical gradient is here and afterward indicated by a dashed circle in all panels.}
    \label{fig:sphere_low-to-high}
\end{figure*}

Figure~\ref{fig:sphere_low-to-high} shows the case $n_\mathrm{1}/n_\mathrm{0} > 1$.
For $n_\mathrm{1}/n_\mathrm{0}=1.2$ with $\vec{r}_{0}=($5\,mm, 5\,mm, 5\,mm), the streamer can propagate through the gradient and enter its interior with multiple irregular branches, see figure~\ref{fig:sphere_low-to-high}(a). 
When the density ratio increases to $n_\mathrm{1}/n_\mathrm{0} \geqslant 1.4$, the streamer is inhibited from propagating through the gradient.
Instead, it propagates along the spherical surface with several branching channels.
After approaching the same height as the sphere's center, the branching channels propagate vertically downwards along the background electric field.
Note that the initial interaction of a streamer with a spherical gradient is similar to that with a planar gradient at the same angle, but that the subsequent propagation depends on the gradient's curvature.

Furthermore, when the spherical gradient is horizontally moved to the right for $n_\mathrm{1}/n_\mathrm{0}=1.4$, as shown in figure~\ref{fig:sphere_low-to-high}(b), the streamers only interact with the left side of the sphere and propagate along its surface.

\begin{figure*}
    \centering
    \includegraphics[width=1\textwidth]{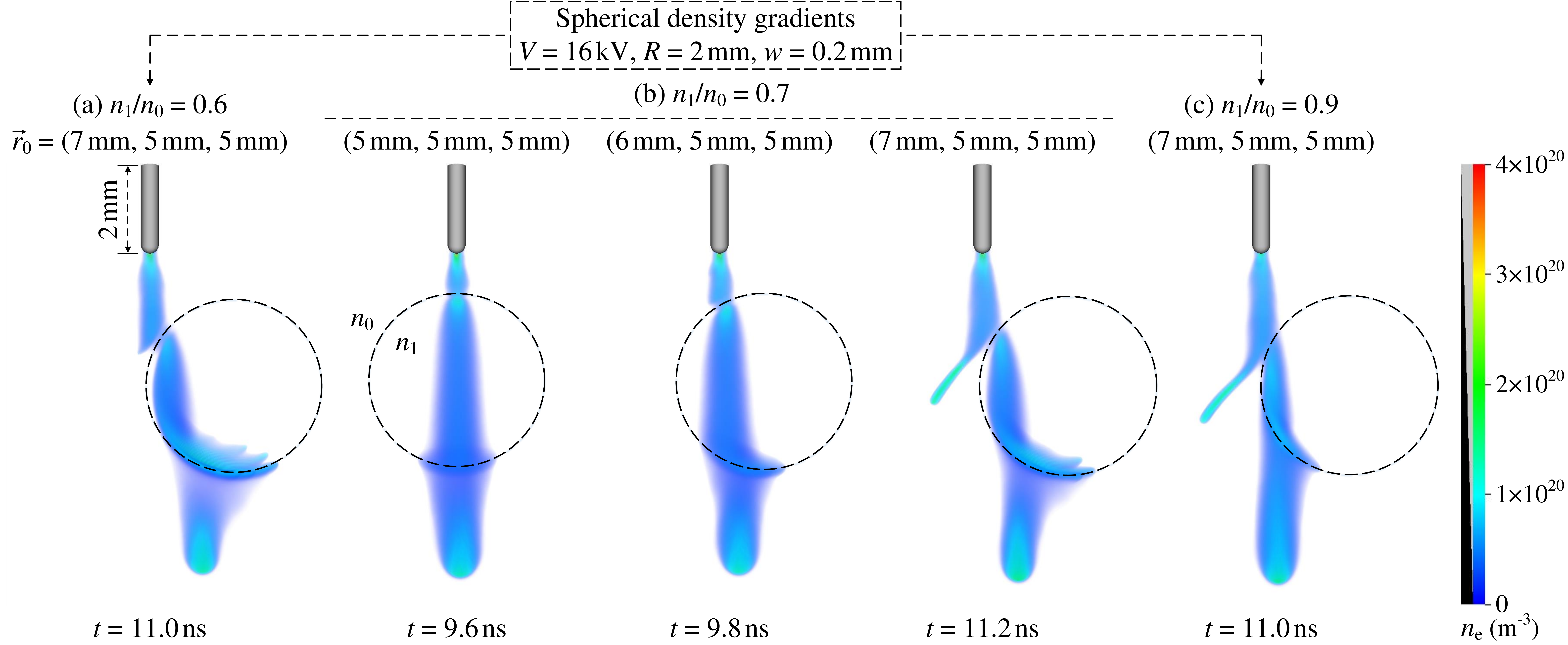}
    \caption{The interaction between a streamer and a spherical density gradient for $n_\mathrm{1}/n_\mathrm{0} < 1$.
    Panels (a) and (c) show streamers for different density ratios $n_\mathrm{1}/n_\mathrm{0}$ at a gradient center $\vec{r}_{0}=($7\,mm, 5\,mm, 5\,mm), whereas panel (b) shows streamers for different gradient centers $\vec{r}_{0}$ at a density ratio $n_\mathrm{1}/n_\mathrm{0}=0.7$.}
    \label{fig:sphere_high-to-low}
\end{figure*}

Figure~\ref{fig:sphere_high-to-low} shows that when $n_\mathrm{1}/n_\mathrm{0} < 1$ the streamers can propagate through the top and bottom of the gradient and form small discontinuities at the gradient.
For $n_\mathrm{1}/n_\mathrm{0}=0.7$ in figure~\ref{fig:sphere_high-to-low}(b), when the spherical gradient is horizontally moved off-center at $\vec{r}_{0}=($7\,mm, 5\,mm, 5\,mm), a branching channel is visible outside the sphere, similar to figure~\ref{fig:plane_high-to-low}(b) for the case $\tan \theta=4$.
Inside the sphere, the discharge splits into two parts upon encountering the bottom of the gradient, where the first part propagates through the gradient and continues propagating in the high-density region, and the second part propagates along the surface.
In addition, increasing the density ratio to $n_\mathrm{1}/n_\mathrm{0}=0.9$ results in less discharge growth inside the sphere, see figure~\ref{fig:sphere_high-to-low}(c).

\subsection{Hemispherical density gradient}\label{sec:hemispherical-gradient}

\begin{figure*}
    \centering
    \includegraphics[width=1\textwidth]{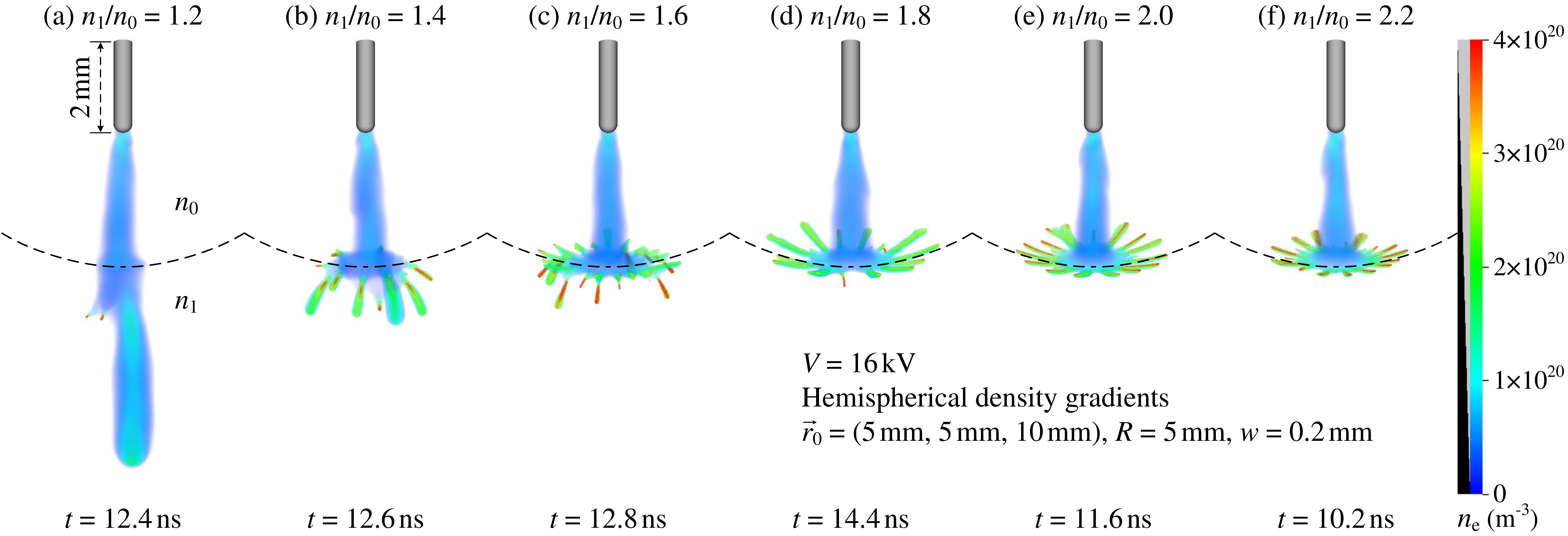}
    \caption{Interaction with a hemispherical density gradient for different density ratios $n_\mathrm{1}/n_\mathrm{0}$.
    The hemispherical gradient is partly indicated by a dashed curve.}
    \label{fig:hemisphere_low-to-high}
\end{figure*}

We now look into the interaction between a positive streamer and a hemispherical density gradient, see figure~\ref{fig:four_gradients}(d) and table~\ref{tab:four-gradients}.
The effect of the air density ratio $n_\mathrm{1}/n_\mathrm{0}$ on streamer propagation from a low-density region to a high-density region is illustrated in figure~\ref{fig:hemisphere_low-to-high}.
The simulations are performed at an applied voltage $V=16$\,kV with a gradient center $\vec{r}_{0}=($5\,mm, 5\,mm, 10\,mm), a radius $R=5$\,mm and a width $w=0.2$\,mm.

Figure~\ref{fig:hemisphere_low-to-high}(a) shows that for $n_\mathrm{1}/n_\mathrm{0}=1.2$ the streamer can propagate through the gradient and continue its propagation with two small branches.
When the density ratio increases to $n_\mathrm{1}/n_\mathrm{0}=1.4$, the streamer splits into several branches when it encounters the gradient, see figure~\ref{fig:hemisphere_low-to-high}(b).
Increasing $n_\mathrm{1}/n_\mathrm{0}$ to fall between 1.6 and 1.8 causes the streamer to split into two parts upon encountering the gradient, where the first part penetrates into the high-density region with multiple branching channels, and the second part propagates sidewards along the surface, see figures~\ref{fig:hemisphere_low-to-high}(c) and (d).
Finally, when $n_\mathrm{1}/n_\mathrm{0}$ exceeds 2.0, the streamer only propagates sidewards along the surface with multiple branching channels, forming a flower-like structure, see figures~\ref{fig:hemisphere_low-to-high}(e) and (f).
These channels slightly bend upwards, which is surprising since the background electric field is directed downwards.
Note that the sideward discharge shows high electron densities in the branching channels, and it decelerates and tends to stagnate.

We find that the ability of a gradient to inhibit streamer propagation depends not only on its angle but also on its shape.
This is demonstrated in figures~\ref{fig:plane_low-to-high_density_ratio},~\ref{fig:sphere_low-to-high} and~\ref{fig:hemisphere_low-to-high}, where we observe that with the same parameters of $V=16$\,kV, $w=0.2$\,mm and $\theta=0$, a convex spherical gradient is the most effective in inhibiting streamer propagation ($n_\mathrm{1}/n_\mathrm{0} \geqslant 1.4$), followed by a planar gradient ($n_\mathrm{1}/n_\mathrm{0} \geqslant 1.8$) and a concave hemispherical gradient ($n_\mathrm{1}/n_\mathrm{0} \geqslant 2.0$).

\subsection{Gradual density variation}\label{sec:gradual-variation}

\begin{figure*}
    \centering
    \includegraphics[width=1\textwidth]{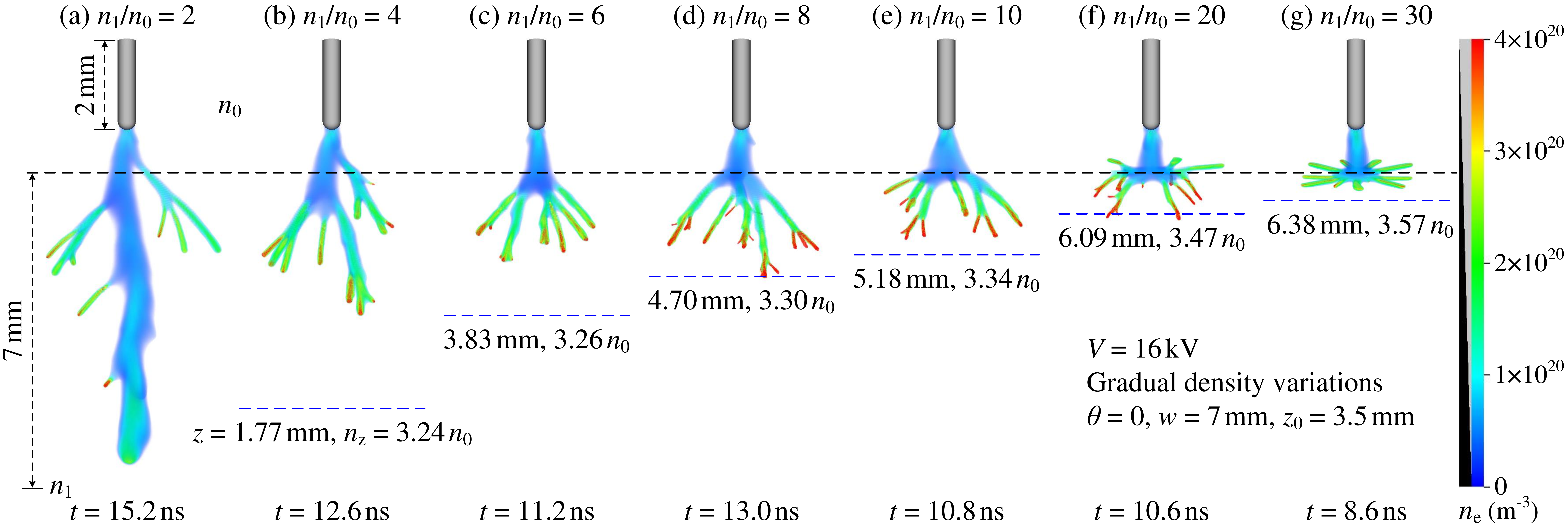}
    \caption{Interaction with gradual density variations for different density ratios $n_\mathrm{1}/n_\mathrm{0}$.
    The gas density linearly increases between the black dashed line and the bottom of the figure, over a width $w = 7 \, \textrm{mm}$.
    For reference, the locations at which the reduced background electric field $E_\mathrm{bg}/n$ is equal to the approximate reduced stability field $E_\mathrm{st} / n_\mathrm{0}$ are indicated by blue dashed lines, using $E_\mathrm{st} = 5 \, \mathrm{kV/cm}$~\cite{allen1991}.
    The height $z$ and the actual gas density $n_\mathrm{z}$ for these blue lines are also given.}
    \label{fig:gradual_low-to-high_density_ratio}
\end{figure*}

Finally, we study how a positive streamer interacts with a much wider density gradient, using $w=7$\,mm, see figure~\ref{fig:four_gradients}(b) and table~\ref{tab:four-gradients}.
The effect of the air density ratio $n_\mathrm{1}/n_\mathrm{0}$ is illustrated in figure~\ref{fig:gradual_low-to-high_density_ratio}.
The simulations are performed at an applied voltage $V=16$\,kV.
We note that they are somewhat unrealistic, as they do not take into account the variation of the photon absorption length with the gas density, which is left for future work.

Figure~\ref{fig:gradual_low-to-high_density_ratio}(a) shows that for $n_\mathrm{1}/n_\mathrm{0}=2$ the streamer can penetrate into the gradient accompanied by the formation of several side branches.
As the density ratio increases, more branching channels form with smaller diameters and higher electron densities at their tips, see figures~\ref{fig:gradual_low-to-high_density_ratio}(b)--(f).
Branching angles increase as well, since it is more difficult for the discharge to propagate downwards inside the gradient.

\begin{figure}
    \centering
    \includegraphics[width=0.48\textwidth]{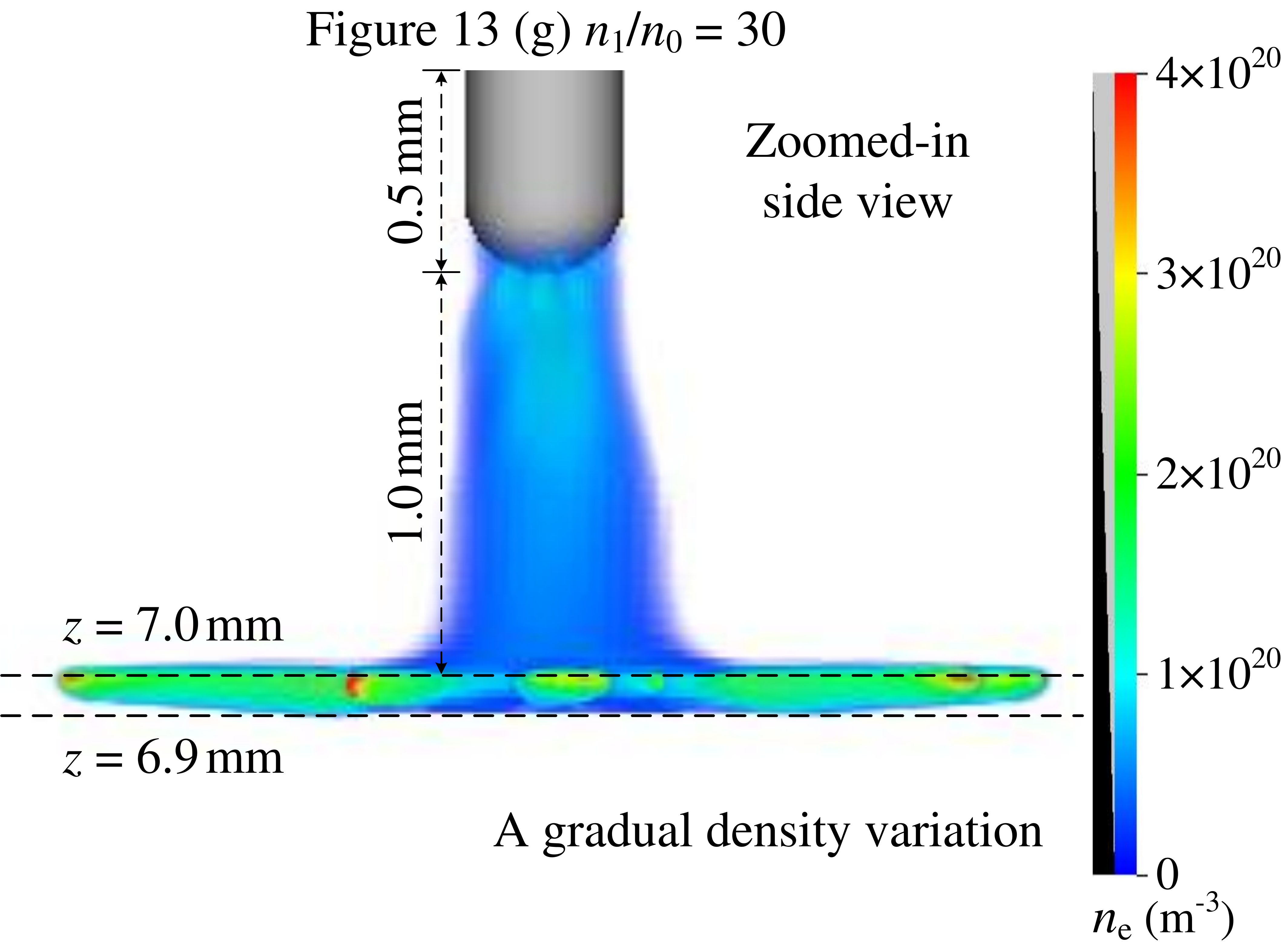}
    \caption{Zoomed-in side view of figure~\ref{fig:gradual_low-to-high_density_ratio}(g) with $n_\mathrm{1}/n_\mathrm{0}=30$.
    The streamer only slightly penetrates into the gradient, over a vertical distance of less than 0.1\,mm.
    Similar behavior was found for other cases in which streamers did not propagate into the high-density region.}
    \label{fig:gradual_dr30_zoom-in}
\end{figure}

When the density ratio increases to $n_\mathrm{1}/n_\mathrm{0}=30$, the streamer is no longer able to penetrate into the gradient but propagates along the surface, see figure~\ref{fig:gradual_low-to-high_density_ratio}(g) and figure~\ref{fig:gradual_dr30_zoom-in}.
We remark that for the other cases streamer propagation is expected to stop after some distance, when the reduced background electric field $E_\mathrm{bg}/n$ becomes too low.
For reference, the locations where $E_\mathrm{bg}/n$ equals the approximate reduced streamer stability field $E_\mathrm{st} / n_\mathrm{0}$ are indicated in figure~\ref{fig:gradual_low-to-high_density_ratio}.

\section{Discussion}\label{sec:discussion}

\subsection{The threshold for inhibiting streamer propagation}\label{sec:plane-threshold}

\begin{figure*}
    \centering
    \includegraphics[width=1\textwidth]{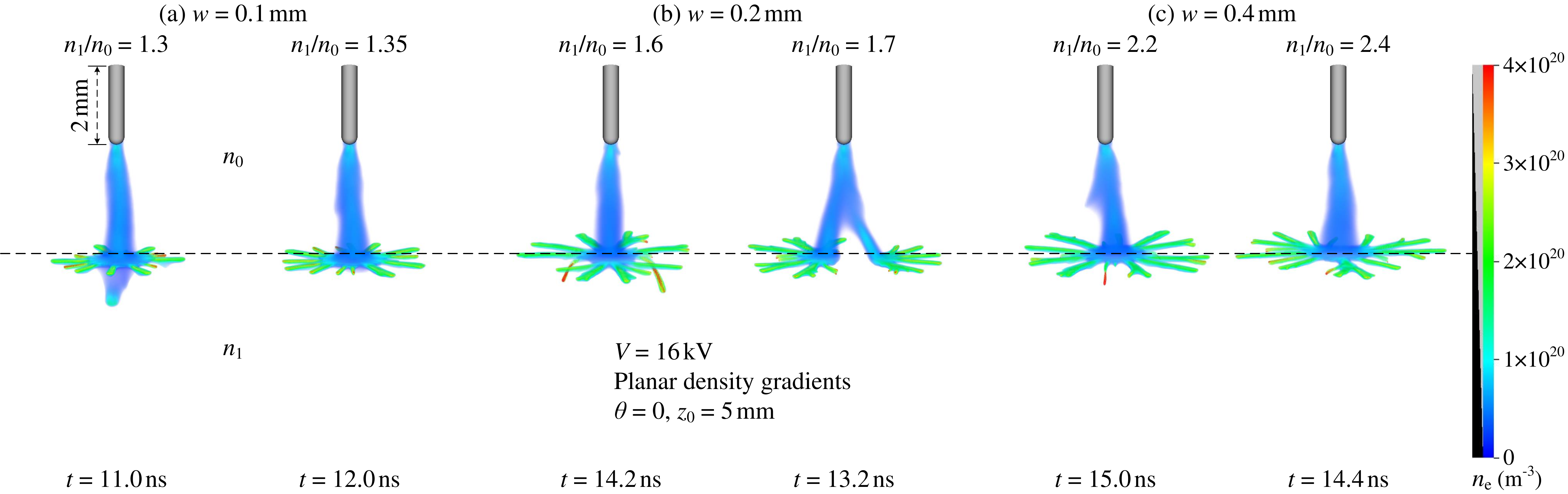}
    \caption{The density ratio $n_\mathrm{1}/n_\mathrm{0}$ required to inhibit streamer propagation through a planar density gradient for three different gradient widths $w$.}
    \label{fig:plane_low-to-high_slope}
\end{figure*}

We have shown that the density ratio threshold for inhibiting streamer propagation from a low-density region to a high-density region depends on the gas gradient width, the type of gradient, the gradient angle and the applied voltage.
To further investigate this threshold, we focus on a planar gradient with a center height $z_\mathrm{0}=5$\,mm and an angle $\theta=0$.
At $V=16$\,kV, we approximately determine the density ratio $n_\mathrm{1}/n_\mathrm{0}$ required to inhibit streamer propagation for various gradient widths, as illustrated in figure~\ref{fig:plane_low-to-high_slope}.
For $w=0.1$\,mm, $w=0.2$\,mm and $w=0.4$\,mm, we find that the required density ratios are about $1.35$, $1.7$ and $2.4$, respectively.

The above values all correspond to a similar density slope, given by
\begin{equation}
  \label{eq:density-slope}
  (n_\mathrm{1} - n_\mathrm{0})/w  \approx 3.5\,n_\mathrm{0}/\mathrm{mm},
\end{equation}
which furthermore agrees rather well with the slope obtained from figure~\ref{fig:gradual_low-to-high_density_ratio}(g), which is about $4 \,n_\mathrm{0}/\mathrm{mm}$.
This can be explained by considering the reduced background electric field $E/n$ ahead of the streamer.
Normally, a streamer grows in the forward direction, since $E/n$ is highest there.
However, when the density slope exceeds a certain threshold, sideward growth is stronger than forward growth (due to the increasing density in the forward direction), which deforms the streamer and leads to propagation along the surface.

We remark that the above density slope for inhibiting streamer propagation is not unique;
it depends on many factors, including the gradient geometry (e.g. its angle) and the primary streamer properties (e.g. the radius), which can be affected by changing the applied voltage and the gas density $n_\mathrm{0}$.
Furthermore, the gradient width should be comparable to the streamer radius or larger.

\subsection{Effect of inhomogeneous gas composition}\label{sec:gas-composition}

In the present study the gas density was spatially varied, leading to a variation in $E/n$ that affected electron transport and reaction coefficients, but the gas composition was always 20\% O$_2$ and 80\% N$_2$.
In some plasma devices, such as plasma jets, there is instead a spatial variation in the gas composition, see e.g.~\cite{naidis2011a, xiong2012, boeuf2013}.
In the different gases, electron transport and reaction coefficients will also differ.
Depending on the particular gases used, such a variation could lead to similar effects as observed in this paper.
However, some of the discharge dynamics can be rather different in such inhomogeneous mixtures, for example due to Penning ionization or due to significant differences in photoionization between the gases.

\subsection{Gas density effect on photoionization}\label{sec:photoionization-effect}

We have not taken the effect of the gas density on the photon absorption length into account when computing photoionization.
These absorption lengths scale like $1/n$~\cite{zheleznyak1982}.
When a positive streamer propagates from a low gas density to a high gas density, this will result in a higher photoionization density at the boundary of the high-density region.
How strong this effect will be depends on the gas density ratio and the width of the gradient, where it should be noted that typical photon absorption distances are rather small (less than a mm in air at 1\,bar).
For sharp gradients and large density ratios, the locally increased photoionization density could further enhance the surface propagation mode, thereby also increasing the ``diode effect'' discussed in~\cite{starikovskiy2019}.

For most cases presented in this paper, the gas density ratio was below a factor of two.
Based on past simulation work~\cite{bagheri2019}, we do not expect major differences in streamer properties like radius or velocity if gas-dependent photon absorption lengths were taken into account.
However, it was recently shown that streamer branching can be quite sensitive to the amount of photoionization~\cite{wang2023}.
We leave the exploration of these effects for future work.

\section{Conclusions}\label{sec:conclusion}

We have studied the effect of gas density inhomogeneities on positive streamer discharges in air using a 3D fluid model, generalizing the 2D axisymmetric fluid simulations of~\cite{starikovskiy2019}.
In order to realistically simulate streamer branching, we included a stochastic photoionization model with discrete photons.
Various types of planar and (hemi)spherical gas density gradients were considered.
Streamers propagated from a region of density $n_\mathrm{0}$ towards a region of higher or lower gas density $n_\mathrm{1}$, where $n_\mathrm{0}$ corresponds to $300\,\mathrm{K}$ and $1\,\mathrm{bar}$.
Streamers could always propagate into a region with a lower gas density and their paths deviated towards nearby low-density regions.
For the case of low-to-high gas density, we observed streamer branching at the density gradient, with branches growing in a flower-like pattern over the gradient surface.
Depending on the gas density ratio, the gradient width and other factors, narrow branches were able to propagate into the higher-density gas.
In a planar geometry, we found such propagation was possible up to a gas density slope of $3.5\,n_\mathrm{0}/\mathrm{mm}$.
This value was dependent on a number of conditions, such as the gradient angle.
Surprisingly, a higher applied voltage made it more difficult for streamers to penetrate into the high-density region, due to an increase of the primary streamer's radius.

\ack
B.G. was funded by the China Scholarship Council (CSC) (Grant No.\,201906280436).

\section*{Data availability statement}
The data that support the findings of this study are openly available at the following URL/DOI: \url{https://doi.org/10.5281/zenodo.7927427}.

\appendix

\section{Implementation of gas density variation}\label{sec:gas-density-expressions}

Below, we explain how the gas density variations were implemented in the code.
For the planar density gradients and gradual density variations, we first define a plane from coefficients $(a, b, c, d)$ as
\begin{equation}
\label{eq:planar-gradients}
    ax + by +cz + d = 0\,.
\end{equation}
For every grid cell in the domain, the signed distance $f(x, y, z)$ to this surface is computed as
\begin{equation}
\label{eq:planar-distance}
    f(x, y, z) = \frac{ax + by +cz + d}{\sqrt{a^2 + b^2 + c^2}}.
\end{equation}
For $f(x, y, z) < -w/2$, where $w$ is the width of the gradient, the gas density $n$ is given by $n = n_0$.
For $f(x, y, z) > w/2$ it is $n = n_1$, and for values in between the density is linearly interpolated between $n_0$ and $n_1$.
When the mesh in refined, the gas density is recomputed on the finer grid.

Spherical and hemispherical density gradients are defined by coordinates $\vec{r}_{0}=(x_0, y_0, z_0)$ and a radius $R$.
The signed distance to the sphere is then determined in every grid cell as
\begin{equation}
\label{eq:spherical-gradients}
    f(x, y, z) = \sqrt{(x - x_0)^2 + (y - y_0)^2 +(z - z_0)^2} - R\,.
\end{equation}
For spherical gradients, we then have $n = n_1$ for $f(x, y, z) < -w/2$ and $n = n_0$ for $f(x, y, z) > w/2$, with densities linearly interpolated in between.
The same is used for hemispherical gradients, but $n_0$ and $n_1$ are swapped, so that the density inside the hemisphere is always $n_0$.

\section*{References}
\normalem
\bibliography{references}

\end{document}